\documentclass[reprint, aip, footinbib,  letterpaper, superscriptaddress]{revtex4-2}

\usepackage[T1]{fontenc} 

\usepackage{amsmath,color, tikz}
\usetikzlibrary{matrix}
\usepackage{amssymb}
\usepackage{amsfonts}
\usepackage{amsthm}
\usepackage{mathtools}

\usepackage{algorithm}
\usepackage{algpseudocode}
\usepackage{dsfont}

\usepackage{bbold}
\usepackage{chemformula}
\usepackage{siunitx}

\DeclarePairedDelimiter\ket{\lvert}{\rangle}
\DeclarePairedDelimiterX\braket[2]{\langle}{\rangle}{#1 \delimsize\vert #2}
\DeclarePairedDelimiterX\braket3[3]{\langle}{\rangle}{#1 \delimsize\vert #2 \delimsize\vert #3}

\usepackage{accents}

\usepackage{fancyvrb}

\newcommand{\vk}{\mathbf{k}}

\newcommand{\vxi}{\boldsymbol{\xi}}

\newcommand{\hH}{\hat{H}}

\newcommand{\avg}[1]{\left\langle #1\right\rangle}

\newcommand{\red}[1]{{\color{black} #1}}

\newcommand{\vkp}{\mathbf{k}_{\parallel} }
\newcommand{\vrp}{\mathbf{r}_{\parallel} }
\newcommand{\kp}{k_{\parallel} }

\begin{document}

	\title{Vibrational Polaritons with Broken In-Plane Translational Symmetry}
	
	\author{Tao E. Li}%
	\email{taoeli@udel.edu}
	\affiliation{Department of Physics and Astronomy, University of Delaware, Newark, Delaware 19716, USA}
	
        \begin{abstract}
        \red{Vibrational polaritons form in a planar Fabry--P\'erot microcavity when a vibrational mode of a layer of molecules is near resonant with an infrared cavity mode. Herein,  dispersion relations of vibrational polaritons are studied when the molecular density distribution breaks the macroscopic translational symmetry along the cavity mirror plane. Both perturbative theory and numerical calculations show that,} if a homogeneous in-plane molecular distribution is modulated by sinusoidal fluctuations, in addition to a pair of upper and lower polariton branches, a discrete number of side polariton branches may emerge in the polariton dispersion relation.  Moreover, for a \red{periodic} Gaussian molecular \red{in-plane} density distribution, only two, yet significantly broadened polariton branches exist in the spectra. This polariton linewidth broadening is \red{caused by} the scattering between cavity modes at \red{neighboring} in-plane frequencies \red{due to the symmetry breaking}, which is distinguished from known origins of polariton broadening such as the homogeneous broadening of molecules, the cavity loss\red{, or the large energetic disorder of molecules}. Associated with the broadened polariton branches, under the \red{periodic} Gaussian in-plane inhomogeneity, a significant number of the VSC eigenstates contain a non-zero contribution from the cavity photon mode at zero in-plane frequency, blurring the distinction between the bright and the dark modes. Looking forward, our theoretical investigation should facilitate the experimental exploration of vibrational polaritons with patterned in-plane molecular density distributions.
	\end{abstract}

	\maketitle

    \section{Introduction}

\begin{figure}
		\centering
		\includegraphics[width=1.0\linewidth]{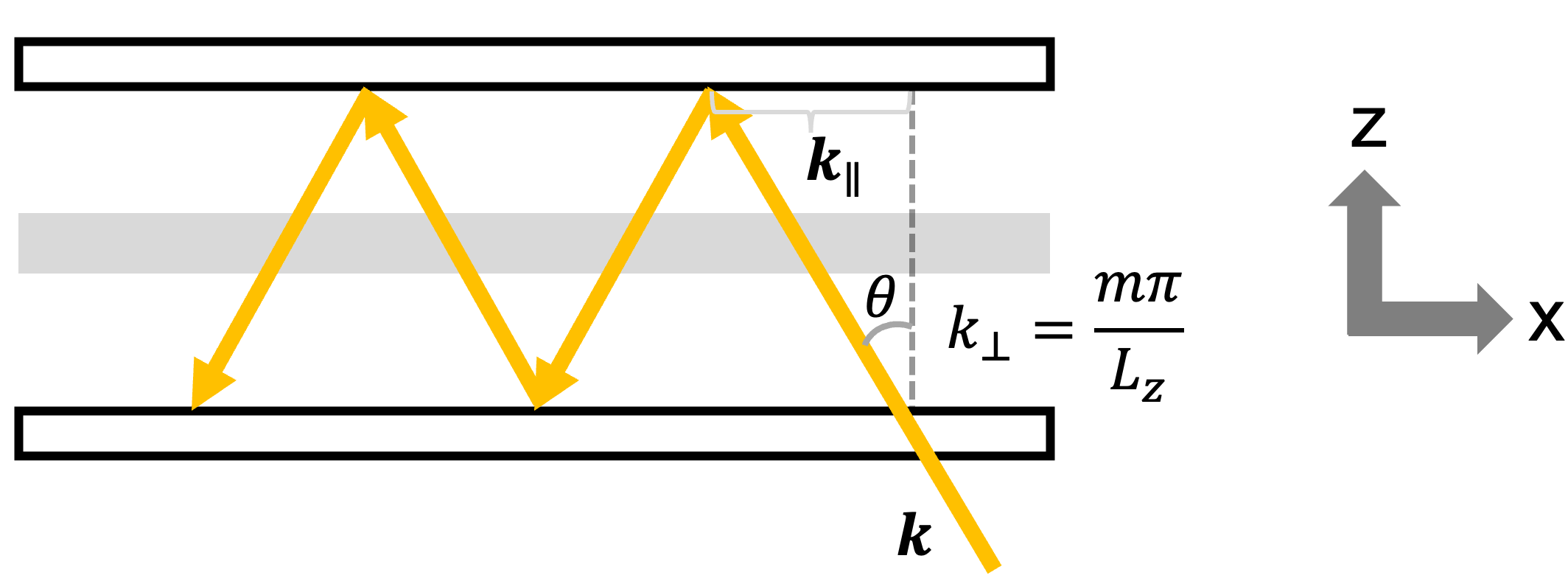}
		\caption{Sketch of a planar Fabry--P\'erot microcavity. In this cavity setup, a pair of parallel cavity mirrors is placed along the $z$-direction with a separation of $L_z$, and the cavity mirrors span over the $xy$-plane. This cavity supports cavity photon modes with the wave vector $\vk = (\vkp, k_{\perp})$, where the in-plane wave vector $\vkp$ is an arbitrary two-dimensional vector in the $xy$-plane, and $k_{\perp} = m\pi/L_z$ with $m=1,2,\cdots$.    A layer of molecules (gray) is placed at the middle of the cavity to form collective strong coupling with the cavity photon modes.
		}
		\label{fig:FP_cavity}
	\end{figure}

    Polaritons, hybrid light-matter states stemming from strong light-matter interactions, have been demonstrated across a wide range of experimental devices \cite{Weisbuch1992,PhysRevLett.113.156401,Shalabney2015,Long2015,Wright2023,Chikkaraddy2016,Damari2019,Canales2021,Yoo2021}. \red{Among different categories of the experimental devices, planar Fabry--P\'erot microcavities have been frequently employed \cite{Deng2010,Carusotto2013}.
    For this planar cavity geometry, as shown in Fig. \ref{fig:FP_cavity}, a layer of molecules is confined between a pair of parallel mirrors. In this cavity,} a continuum of photon modes with the wave vector $\vk = (\vkp, k_{\perp})$ is supported, where $\vkp$ denotes an arbitrary in-plane wave vector oriented along the cavity mirror plane, and the discrete perpendicular wave vector $k_{\perp} = m\pi/L_z$ (with $m = 1,2,\cdots$) is determined by the cavity length $L_z$. 

    When theoretical models are used to describe the polariton dispersion relation (i.e., polariton spectra as a function of $\vkp$) in a planar Fabry--P\'erot microcavity, \red{the cavity modes at different $\vkp$ values are often assumed to be independent with each other.\cite{Deng2010} With this independent-mode approximation, the polariton dispersion relation is  evaluated by  calculating the polariton eigenstates between the molecules and the cavity mode at each individual $\vkp$ value independently \cite{Deng2010}. Throughout this manuscript, we will also refer this independent-mode approximation as the single-mode approximation.}

    The validity of the single-mode approximation stems from the translational symmetry of the molecules along the cavity mirror plane \cite{Agranovich2003}. With this molecular in-plane translational symmetry, \red{during the calculation of  the polariton dispersion relation,} the scattering events between cavity modes at different $\vkp$ values cancel with each other \cite{Agranovich2003}. As a result, the single-mode approximation becomes valid. \red{With the importance of in-plane translational symmetry in mind, a nature question is, can the breakdown of this symmetry affect the polariton dispersion relation in a planar Fabry--P\'erot microcavity? 
    
    Generally speaking, the in-plane translational symmetry breaking can  manifest in two length scales: the microscopic (with a length scale much smaller than the wavelength of the cavity photons), and the macroscopic (with a length scale equivalent or larger than the wavelength of the cavity photons). Microscopically, the in-plane translational symmetry breaking can stem from factors such as the position and orientation disorders of the molecules \cite{Agranovich2003,Michetti2005,Litinskaya2006}, or the use of low-symmetry crystals \cite{Galiffi2023} in the cavity. In this microscopic length scale, the impact of in-plane translational symmetry breaking on the polariton dispersion relation has been widely studied. It has been found that the coherence length of exciton-polaritons  can be significantly reduced due to molecular disorders \cite{Agranovich2003,Michetti2005,Litinskaya2006}. 

    Macroscopically, the molecular in-plane translational symmetry could be broken by engineering  the in-plane molecular density distribution. This includes applying  a local strain to the exciton layer or preparing artificial lattice structures of the materials  (with a lattice constant of a few $\mu$m) 
    \cite{Tanese2013,Jacqmin2014,Schneider2017,Baranov2020}. These techniques can facilitate the formation of exciton-polariton condensates  by creating trapping potentials.\cite{Tanese2013,Schneider2017} Beyond the scope of exciton-polaritons in optical cavities, preparing the patterned surfaces also provides a means to control the band structure of plasmon-exciton polaritons \cite{Chevrier2019}.
    }

    

    
    \red{For vibrational strong coupling (VSC) observed in the past decade \cite{Shalabney2015,Long2015}, the study of symmetry breaking is limited.} Due to the potential of modifying chemical reaction rates and energy transfer pathways, VSC has attracted great attention both experimentally \cite{Thomas2016, Ahn2023Science,Thomas2019_science,Wright2023,Hirai2020Crys,Imperatore2021,Wiesehan2021,Fidler2023,Dunkelberger2016,Xiang2020Science,Chen2022,Simpkins2023} and theoretically \cite{Galego2019,Campos-Gonzalez-Angulo2019,Hoffmann2020,LiHuo2021,Fischer2021,YangCao2021,Wang2022JPCL,Flick2017,Riso2022,Schafer2021,Bonini2021,Yang2021,Rosenzweig2022,Triana2020Shape,Li2020Water,Campos-Gonzalez-Angulo2023,Suyabatmaz2023,Ribeiro2018,Li2022Review,Fregoni2022,Mandal2023ChemRev,Ruggenthaler2023}. In this field, Xiang \textit{et al} experimentally reported the preparation of periodic cavity patterns with a size of 50 $\mu$m along the cavity mirror plane   and studied the nonlinear interactions between vibrational polaritons in this uneven cavity mirror structure \cite{Xiang2021}. 
    \red{More recently, Suyabatmaz and Ribeiro\cite{Suyabatmaz2023} theoretically investigated the transport behavior of vibrational polaritons due to microscopic molecular disorders.}

    \red{Here, we theoretically study the polariton dispersion relation under VSC when the molecular density distribution exhibits large-scale, macroscopic in-plane translational symmetry breaking. The necessity of this study is twofold. First,} the macroscopic in-plane molecular  inhomogeneity might be pervasive in VSC experiments.  For example, among VSC experiments on thermally activated chemical reactions \cite{Thomas2016,Thomas2019_science,Ahn2023Science,Imperatore2021,Wiesehan2021,Fidler2023} and crystallization processes \cite{Hirai2020Crys},  the large surface tension at the cavity-molecular micro-interface may \red{potentially} prevent molecular distributions along the cavity mirror plane from maintaining \red{large-scale, macroscopic} homogeneity during the experiments.
    \red{Second, under VSC, the typical photon wavelength corresponding to molecular vibrations is an order of magnitude larger than that of exciton-polaritons, making the experimental fabrication of VSC devices with large-scale molecular density inhomogeneity potentially more convenient than those under electronic strong coupling. This study is also inspired by} the numerical evidence from cavity molecular dynamics simulations, which suggest a correlation between the in-plane translational symmetry breaking and the complexity in the polariton dispersion relation \cite{Li2024CavMD}.
    Therefore, it is crucial to understand in detail how the in-plane molecular distribution inhomogeneity, especially those distributions beyond simple lattice forms, may impact the spectra and dynamical processes of vibrational polaritons. 
        
    In this manuscript, we provide an analytical 1D solution of the polariton dispersion relation when a homogeneous in-plane molecular density distribution is modulated by weak sinusoidal fluctuations. The analytical solution is then generalized to the case of an arbitrary weak in-plane molecular density inhomogeneity. Moreover, beyond the perturbative limit, we perform numerical calculations to study the polariton dispersion relations for a few in-plane molecular distributions in both 1D and two dimensions (2D). 
    
    While it is known that a homogeneous in-plane molecular distribution leads to pair of polariton branches in the dispersion relation [i.e., the upper polariton (UP) and the lower polariton (LP) branches] \cite{Deng2010,Carusotto2013}, our calculations show that the sinusoidal inhomogeneity in the molecular density distribution generates a few additional side polariton branches in the spectra. More interestingly, a \red{periodic} Gaussian in-plane molecular distribution generates only two, yet significantly broadened polariton branches in the dispersion relation. Such a polariton broadening is not due to the well-known origins of polariton broadening such as the homogeneous broadening of molecules\cite{Houdre1996,Long2015}, the cavity loss\red{, or the large energy disorder of molecules\cite{George2024}.} Instead, it stems from the \red{breakdown of large-scale in-plane translational symmetry} and the complicated scattering events between cavity modes at different $\vkp$ values. In the case of the \red{periodic} Gaussian in-plane density distribution, the distinction between the bright and the dark modes is also blurred, which may impact various dynamical processes under VSC.

    This paper is organized as follows. Sec. \ref{sec:theory} presents the analytical theory of microcavity polaritons with broken in-plane translational symmetry and the perturbative calculations. Sec. \ref{sec:details} provides details on the numerical calculations. Sec. \ref{sec:results} shows the polariton dispersion relations for a few 1D and 2D in-plane molecular density distributions. Sec. \ref{sec:discussion} analyzes the photonic weight distribution among VSC eigenstates. We conclude in Sec. \ref{sec:conclusion}.

    \section{Theory}\label{sec:theory}

    \subsection{The single-molecule single-mode limit: Jaynes--Cummings model}
    We start with perhaps the simplest theoretical description of polaritons, the Jaynes--Cummings (JC) model \cite{Jaynes1963} under the rotating wave approximation. In this model, a single molecule with transition frequency $\omega_0$ is coupled to a single cavity photon with frequency $\omega_{\rm c}$:
    \begin{equation}
    \label{eq:JC_Hamiltonian}
        \hH_{\rm JC} = \omega_0 \hat{b}^{\dagger} \hat{b} + \omega_{\rm c} \hat{a}^{\dagger} \hat{a} + g_0 \left ( \hat{a}^{\dagger}\hat{b} + \hat{a}\hat{b}^{\dagger} \right ).
    \end{equation}
    Here, $\hat{b}^{\dagger}$ ($\hat{a}^{\dagger}$) and $\hat{b}$ ($\hat{a}$) denote the creation and the annihilation operator of the molecule (cavity photon), respectively, and $g_0$ denotes the light-matter coupling strength. In the original version of the JC model \cite{Jaynes1963}, the molecule was represented by a two-level system, but here the molecule is represented by a quantum harmonic oscillator to better accommodate the situation of VSC.
    
    In the single-excitation manifold, the eigen equation of the above Hamiltonian reads:
    \begin{equation}
    \label{eq:JC_eigen_equation}
        \begin{pmatrix}
            \omega_0 & g_0 \\
            g_0 & \omega_{\rm c}
        \end{pmatrix}
        \begin{pmatrix}
            \beta \\
            \alpha
        \end{pmatrix}
        =
        \omega 
        \begin{pmatrix}
            \beta \\
            \alpha
        \end{pmatrix}.
    \end{equation}
    In Eq. \eqref{eq:JC_eigen_equation}, the corresponding eigenvalues $\omega$ can be obtained by solving
    \begin{equation}\label{eq:JC_eigen_equation_simplified}
        (\omega_{0} - \omega) (\omega_{\rm c} - \omega) = g_0^2.
    \end{equation}
    The resulting two eigen energies are $\omega = \omega_{\pm}$, where
    \begin{equation}
        \omega_{\pm} = \frac{\omega_{0} + \omega_{\rm c}}{2} \pm \sqrt{g_0^2 + \frac{(\omega_0 - \omega_{\rm c})^2}{4}}.
    \end{equation}
    At resonance ($\omega_0 = \omega_{\rm c}$), the Rabi splitting between the two eigen energies is $\Omega \equiv \omega_{+} - \omega_{-} = 2g_0$. Conventionally speaking, polaritons form when the experimentally observed Rabi splitting ($\Omega$) is greater than the linewidth of either the molecular or the photonic transition linewidth. In each of the polariton state, the molecular weight is
    \begin{equation}
        \left | \beta^{\pm} \right |^2 = \frac{g_0^2}{g_0^2 + (\omega_0 - \omega_{\pm})^2},
    \end{equation}
    and the photonic weight  is
    \begin{equation}
        \left | \alpha^{\pm} \right |^2 = 1- \left | \beta^{\pm} \right |^2 =  \frac{(\omega_0 - \omega_{\pm})^2}{g_0^2 + (\omega_0 - \omega_{\pm})^2} .
    \end{equation}

    \subsection{Many molecules in a Fabry--P\'erot microcavity: Extended Tavis--Cummings model}

    The JC model is usually adequate to predict polariton energies in the strong coupling limit. When optical cavities are used, however, because the light-matter coupling for a single molecule ($g_0$) is negligibly small compared with the molecular or the photonic linewidth, polariton formation often requires a large collection of molecules confined in the cavity. In this collective strong coupling limit, the Tavis--Cummings (TC) model is frequently used \cite{Tavis1968,Tavis1969}:
    \begin{equation}
    \label{eq:TC_Hamiltonian}
        \hH_{\rm TC} = \sum_{j=1}^{N}\omega_0 \hat{b}^{\dagger}_j \hat{b}_j + \omega_{\rm c} \hat{a}^{\dagger} \hat{a} + \sum_{j=1}^{N} g_0 \left ( \hat{a}^{\dagger}\hat{b}_j + \hat{a}\hat{b}^{\dagger}_j \right ).
    \end{equation}
    Here, the molecular Hamiltonian is represented by a collection of $N$ harmonic oscillators instead of a single harmonic oscillator as in the JC model, while the photonic part is still represented by a single harmonic oscillator.

    Below, we are interested in the experimental setup of a layer of molecules confined in a planar Fabry--P\'erot microcavity, as shown in Fig. \ref{fig:FP_cavity}. For this setup, because the Fabry--P\'erot microcavity can support many cavity modes, one may question the validity of the single-mode approximation in the TC Hamiltonian. To this end, we will explicitly include many cavity modes in the Hamiltonian, using the following extended Tavis--Cummings Hamiltonian \cite{Agranovich2003,Michetti2005}:
    \begin{subequations}
    \label{eq:extened_TC_Hamiltonian}
        \begin{equation}
        \label{eq:extened_TC_Hamiltonian_overall}
        \hH_{\rm eTC} = \hH_{\rm M} + \hH_{\rm ph} + \hH_{\rm I} .
    \end{equation}
    Here, the same as the TC Hamiltonian, the molecular Hamiltonian is represented by a collection of $N$ identical harmonic oscillators:
    \begin{equation}
        \hH_{\rm M} = \sum_{j=1}^{N}\omega_0 \hat{b}^{\dagger}_j \hat{b}_j.
    \end{equation}
    Different from the TC Hamiltonian, the photonic Hamiltonian contains all the fundamental cavity modes supported by the Fabry--P\'erot cavity:
    \begin{equation}
    \label{eq:Hph_eTC}
        \hH_{\rm ph} = \sum_{\vkp} \omega_{\rm c}(\vkp)\hat{a}^{\dagger}_{\vkp}\hat{a}_{\vkp}.
    \end{equation}
    In Eq. \eqref{eq:Hph_eTC},  $\vkp$ denotes the in-plane wave vector of the supported photon modes, and the corresponding photonic frequency reads
    \begin{equation}\label{eq:cavity_freq}
        \omega_{\rm c}(\vkp) = c\sqrt{\left( \frac{\pi}{L_{z}} \right )^2 + |\vkp|^2}.
    \end{equation}
    Here, $c = c_0/n_{\rm ref}$ denotes the speed of light in the medium, where $c_0$ denotes the speed of light in the vacuum and $n_{\rm ref}$ denotes the refractive index of the medium; $L_z$ denotes the separation between the two parallel cavity mirrors (or the length of the cavity). Because we are interested in only the fundamental cavity modes, in the above equation the perpendicular wave vector takes $k_{\perp} = \frac{\pi}{L_{z}}$ instead of $\frac{m\pi}{L_{z}}$ ($m=1,2,\cdots$). \red{This simplification rules out the possibility of describing multiple cavity modes at different $m$ values interacting with the molecular or material excitations.\cite{Baranov2020}}
    Finally, in Eq. \eqref{eq:extened_TC_Hamiltonian_overall}, the light-matter coupling Hamiltonian reads:
    \begin{equation}\label{eq:HI_eTC}
        \hH_{\rm I} = \sum_{j=1}^{N} \sum_{\vkp} g_0 e^{i \vkp \cdot \vrp^j} \hat{a}^{\dagger}_{\vkp} \hat{b}_j + \text{h.c.},
    \end{equation}
    where $e^{i \vkp \cdot \vrp}$ denotes the phase of each photon mode at location $\vrp^j$, the in-plane position of molecule $j$, and \text{h.c.} denotes the Hermitian conjugate. In Eq. \eqref{eq:HI_eTC}, only the spatial \red{variation} along the cavity mirror plane is included (the $e^{i \vkp \cdot \vrp^j}$ term),  while the spatial variance perpendicularly to the cavity mirror plane (the $z$-direction in Fig. \ref{fig:FP_cavity}) is neglected. Such a simplification is valid when a layer of molecule is placed at the middle of the cavity, which is our assumption (see also Fig. \ref{fig:FP_cavity}). \red{Here, the molecular layer is assumed to span along the infinite cavity mirror plane, although periodic boundary conditions along the cavity mirror plane will be applied during numerical calculations in Sec. \ref{sec:details}. In Eq. \eqref{eq:HI_eTC}, we also assume the light-matter coupling $g_0$ to be  a constant and neglect its weak $\vkp$ dependence\cite{Agranovich2003}. }
    \end{subequations}

    In the single-excitation manifold $\{ \hat{b}_j^{\dagger}\ket{0}, \hat{a}_{\vkp}^{\dagger}\ket{0}; \forall j, \vkp \}$, the extended TC Hamiltonian reads

\begin{widetext}
\begin{equation}
\hH_{\rm eTC} = 
\begin{tikzpicture}[baseline=(current bounding box.center)]
\matrix (m) [matrix of math nodes,nodes in empty cells,right delimiter={)},left delimiter={(} ]{
\omega_0  & \dots & 0  & g_0 e^{i\vkp\cdot \vrp^j} & \dots & g_0e^{i\vkp'\cdot \vrp^j}   \\
\vdots  & \ddots & \vdots & \vdots & \ddots & \vdots  \\
0 & \dots & \omega_0 & g_0 e^{i\vkp \cdot \vrp^{j'}} & \dots & g_0 e^{i\vkp'\cdot \vrp^{j'}}   \\
g_0 e^{-i\vkp\cdot \vrp^{j}} & \dots & g_0 e^{-i\vkp\cdot \vrp^{j'}}  & \omega_{\rm c}(\vkp) &  \dots &  0 \\
\vdots & \ddots & \vdots & \vdots & \ddots &  \vdots  \\
g_0 e^{-i\vkp' \cdot \vrp^{j}} & \dots & g_0 e^{-i\vkp'\cdot \vrp^{j'}} & 0 & \dots &  \omega_{\rm c}(\vkp')\\
} ;
\end{tikzpicture} .
\label{eq:H_eTC_matrix}
\end{equation}
\end{widetext}
Assuming that the eigenvectors of this extended TC Hamiltonian take the form of $\vxi = (\beta_j \dots \beta_{j'} \   \alpha_{\vkp} \dots  \alpha_{\vkp'})^T$, we can solve the eigen equation of the extended TC Hamiltonian as $\hH_{\rm eTC} \vxi = \omega \vxi$, or
\begin{widetext}
\begin{equation}
\begin{tikzpicture}[baseline=(current bounding box.center)]
\matrix (m) [matrix of math nodes,nodes in empty cells,right delimiter={)},left delimiter={(} ]{
\omega_0 - \omega  & \dots & 0  & g_0 e^{i\vkp\cdot \vrp^j} & \dots & g_0e^{i\vkp'\cdot \vrp^j}   \\
\vdots  & \ddots & \vdots & \vdots & \ddots & \vdots  \\
0 & \dots & \omega_0 - \omega & g_0 e^{i\vkp \cdot \vrp^{j'}} & \dots & g_0 e^{i\vkp'\cdot \vrp^{j'}}   \\
g_0 e^{-i\vkp\cdot \vrp^{j}} & \dots & g_0 e^{-i\vkp\cdot \vrp^{j'}}  & \omega_{\rm c}(\vkp) - \omega &  \dots &  0 \\
\vdots & \ddots & \vdots & \vdots & \ddots &  \vdots  \\
g_0 e^{-i\vkp' \cdot \vrp^{j}} & \dots & g_0 e^{-i\vkp'\cdot \vrp^{j'}} & 0 & \dots &  \omega_{\rm c}(\vkp') - \omega\\
} ;
\end{tikzpicture}
\begin{pmatrix}
        \beta_j \\ \vdots \\ \beta_{j'} \\ \alpha_{\vkp} \\ \vdots \\ \alpha_{\vkp'}
    \end{pmatrix}  = 0 .
\label{eq:H_eTC_eigen_matrix}
\end{equation}
\end{widetext}
Equivalently, the following set of equations is needed to be solved:
\begin{subequations}
    \begin{align}
    \label{eq:eHC_equation_1}
        (\omega_0 - \omega) \beta_j + \sum_{\vkp'}g_0e^{i\vkp'\cdot \vrp^j} \alpha_{\vkp'} & = 0 ,  \\
        \label{eq:eHC_equation_2}
        \sum_{j} g_0 e^{-i\vkp \cdot \vrp^j} \beta_j + \left [ \omega_{\rm c}(\vkp) - \omega\right ]\alpha_{\vkp} &= 0.
    \end{align}
\end{subequations}
In Eq. \eqref{eq:eHC_equation_1}, the index $j$ runs over $j=1,2\cdots, N$; in Eq. \eqref{eq:eHC_equation_2}, the index $\vkp$ runs over all the supported values in the cavity.
By substituting Eq. \eqref{eq:eHC_equation_1}, or $\beta_j = \sum_{\vkp'}g_0e^{i\vkp'\cdot \vrp^j} \alpha_{\vkp'} / (\omega - \omega_0) $, into Eq. \eqref{eq:eHC_equation_2}, we obtain the equations containing only the photonic vector coefficients $\alpha_{\vkp}$:
\begin{equation}\label{eq:eTC_combined_eign_equation}
    \left [ \omega - \omega_{\rm c}(\vkp) \right ] (\omega - \omega_0)\alpha_{\vkp} = \sum_j\sum_{\vkp'} g_0^2 e^{-i(\vkp - \vkp')\vrp^j} \alpha_{\vkp'} .
\end{equation}
Here, the index $\vkp$ runs over all the supported values in the cavity. Solving this large set of equations ($\forall \alpha_{\vkp}$)  yields all the eigenstates of the extended TC Hamiltonian. A similar form of Eq. \eqref{eq:eTC_combined_eign_equation} has been obtained by Agranovich \textit{et al} \cite{Agranovich2003}.

\subsection{A homogeneous molecular distribution: Recovering the JC solution}
Now, let us assume that the molecular distribution is homogeneous along the cavity mirror plane. In this homogeneous limit, following Agranovich \textit{et al} \cite{Agranovich2003},  we can replace the summation over molecules $\sum_j$ in Eq. \eqref{eq:eTC_combined_eign_equation} by an integral $\int \frac{N}{S} d\vrp$, where $S$ denotes the area of the cavity mirror plane and $N/S$ is the density of the molecular distribution along the cavity mirror plane. With this replacement, Eq. \eqref{eq:eTC_combined_eign_equation} becomes
\begin{equation}\label{eq:combined_eq_homo}
\begin{aligned}
    & \left [ \omega - \omega_{\rm c}(\vkp) \right ] (\omega - \omega_0)\alpha_{\vkp} \\
    = & \ \frac{N}{S} \sum_{\vkp'} g_0^2 \int d\vrp e^{-i(\vkp - \vkp')\vrp} \alpha_{\vkp'} .
\end{aligned}
\end{equation}
Now, we invoke the identity $\frac{1}{(2\pi)^{d/2}} \int d\vrp e^{-i(\vkp - \vkp')\vrp}  = \delta(\vkp - \vkp')$, where $d$ denotes the dimension of the cavity mirror plane ($d=1$ or 2). Moreover, we further replace the summation $\sum_{\vkp'} $ by an integral $\int f(\vkp') d\vkp'$, where $f(\vkp')$ denotes the in-plane photonic density of states. With these considerations in mind, at the right hand side of Eq. \eqref{eq:combined_eq_homo}, all the terms with indexes $\vkp' \neq \vkp$ vanish, and Eq. \eqref{eq:combined_eq_homo} can be reduced to
\begin{equation}\label{eq:definition_of_Delta}
    \left [ \omega - \omega_{\rm c}(\vkp) \right ] (\omega - \omega_0) = \frac{(2\pi)^{d/2} N}{S} g_0^2 f(\vkp) \equiv \Delta^2 .
\end{equation}
This equation is identical to the eigen equation for the JC model [Eq. \eqref{eq:JC_eigen_equation_simplified}]. The resulting polariton dispersion relation is 
\begin{equation}
\label{eq:polariton_dispersion_relation}
        \omega_{\pm}(\vkp) = \frac{\omega_{0} + \omega_{\rm c}(\vkp)}{2} \pm \sqrt{\Delta^2 + \frac{[\omega_0 - \omega_{\rm c} (\vkp)]^2}{4}}.
\end{equation}
where the cavity frequency $\omega_{\rm c}(\vkp)$ has been defined in Eq. \eqref{eq:cavity_freq}.
When $\omega_{\rm c}(\vkp) = \omega_0$, the  Rabi splitting 
\begin{equation}
    \Omega_{N} = 2\Delta \propto \sqrt{N} g_0
\end{equation}
is proportional to $\sqrt{N}$. This is the well-known result of the collective Rabi splitting in the conventional TC Hamiltonian \cite{Tavis1968,Tavis1969}.

It is clear that in Eq. \eqref{eq:eTC_combined_eign_equation},  cavity modes at different $\vkp $ values may interact with each other. Only in the limit of a homogeneous molecular distribution along the cavity mirror plane, we can replace the summation \red{$\sum_j e^{-i(\vkp - \vkp')\vrp^j}$ by the identity $\frac{N}{S} \int d\vrp e^{-i(\vkp - \vkp')\vrp}  = \frac{N}{S}(2\pi)^{d/2}\delta(\vkp - \vkp')$, and then cancel out all the scattering events between the cavity modes at $\vkp' \neq \vkp$ values \cite{Agranovich2003}. As a result, polaritons at different $\vkp$ values do not interact with each other, which constitutes the single-mode approximation. Without a homogeneous molecular distribution, $\sum_j e^{-i(\vkp - \vkp')\vrp^j} = \int d\vrp \rho(\vrp)e^{-i(\vkp - \vkp')\vrp}$, where the molecular in-plane density $\rho(\vrp)$ is not a constant. Therefore, we cannot replace $\sum_j e^{-i(\vkp - \vkp')\vrp^j}$ by a simple delta function and cancel out the scattering events at $\vkp' \neq \vkp$.}  Formally speaking, the in-plane translational symmetry of molecules validates the single-mode approximation and greatly simplifies the polariton dispersion relation in planar Fabry--P\'erot microcavities.

\subsection{A molecular distribution with small inhomogeneity: Perturbative treatments}

While the above derivations have been shown in the literature \cite{Agranovich2003},  in this manuscript, we are interested in the question that how an inhomogeneous molecular distribution invalidates the single-mode approximation and reshapes the polariton  dispersion relation in planar Fabry--P\'erot microcavities. 

Generally speaking, for an arbitrary inhomogeneous molecular distribution, finding an analytical solution of the polariton  dispersion relation is very challenging. However, in the limit of a small in-plane density inhomogeneity, because the in-plane translational symmetry is not completely broken, it is still possible to find an analytical solution of the polariton  dispersion relation with perturbative treatments. 

For example, let us assume that the molecular density distribution along the cavity mirror plane takes the following form:
\begin{equation}\label{eq:rho_perturb}
    \rho(\vrp) = \rho_0 + \delta \rho_1(\vrp) ,
\end{equation}
where $\rho_0  = N/S$ denotes the density of a homogeneous molecular distribution,  $\rho_1(\vrp)$ denotes the density inhomogeneity, and $\delta \rightarrow 0$ is a small dimensionless variable to control the perturbative expansion. With Eq. \eqref{eq:rho_perturb}, we can replace the summation $\sum_j$ in Eq. \eqref{eq:eTC_combined_eign_equation} by $\int \rho(\vrp) d\vrp = \int \rho_0 d\vrp + \delta \int \rho_1(\vrp) d\vrp$. As a result, Eq. \eqref{eq:eTC_combined_eign_equation} becomes
\begin{equation}
\label{eq:inhomo_eigen_rho1}
    \begin{aligned}
        & \left [ \omega - \omega_{\rm c}(\vkp) \right ] (\omega - \omega_0)\alpha_{\vkp} \\
        = & \ \Delta^2 \alpha_{\vkp} + (2\pi)^{d/2}\sum_{\vkp'} \delta g_0^2  \widetilde{\rho}_1(\vkp - \vkp')  \alpha_{\vkp'} ,
    \end{aligned}
\end{equation}
where $\Delta$ has been defined in Eq. \eqref{eq:definition_of_Delta}, and  $\widetilde{\rho}_1(\vkp - \vkp') = \frac{1}{(2\pi)^{d/2}}\int  d\vrp \rho_1(\vrp) e^{-i(\vkp - \vkp')\vrp}$ denotes the density inhomogeneity in $k$-space.

\subsubsection{1D sinusoidal inhomogeneity}

To proceed, we now assume that the cavity mirror plane is 1D ($d=1$), and the density inhomogeneity takes the following analytical form:
\begin{equation}\label{eq:sin_density}
    \bar{\rho}_1(x) = \frac{S}{N}\rho_1(x) =  \sin(k_x x),
\end{equation}
where the cavity mirror plane is assumed to span along the $x$ axis, and $\bar{\rho}_1(x)$ denotes the dimensionless inhomogeneity.
Because the Fourier transform of $\rho_1(x)$ in $k$-space is $\widetilde{\rho}_1(\kp - \kp') = \frac{N}{S}\frac{1}{2i}\left[ \delta(\kp - \kp'-k_x) - \delta(\kp - \kp'+k_x) \right]$, Eq. \eqref{eq:inhomo_eigen_rho1} can be reduced to
\begin{equation}
\label{eq:inhomo_eigen_simplified}
    \begin{aligned}
        & \left [ \omega - \omega_{\rm c}(\kp) \right ] (\omega - \omega_0)\alpha_{\kp} \\
         = & \ \Delta^2 \alpha_{\kp} +  \frac{1}{2i}\delta \Delta^2 \left( \alpha_{\kp - k_x} - \alpha_{\kp + k_x} \right) ,
    \end{aligned}
\end{equation}
\red{where $\kp \in (-\infty, +\infty)$ denotes the in-plane wave vector in 1D.}
In this simplified form, the cavity mode at $\kp$ interacts only with two cavity modes at $\kp \pm k_x$. \red{Note that including both the positive and negative values of $\kp$ is important for our calculations of the polariton dispersion relation. In fact, it has been reported that including both the positive and negative $\kp$ values can significantly influence the polariton transport behavior in 1D.\cite{Aroeira2023}}

\begin{figure}
		\centering
		\includegraphics[width=1.0\linewidth]{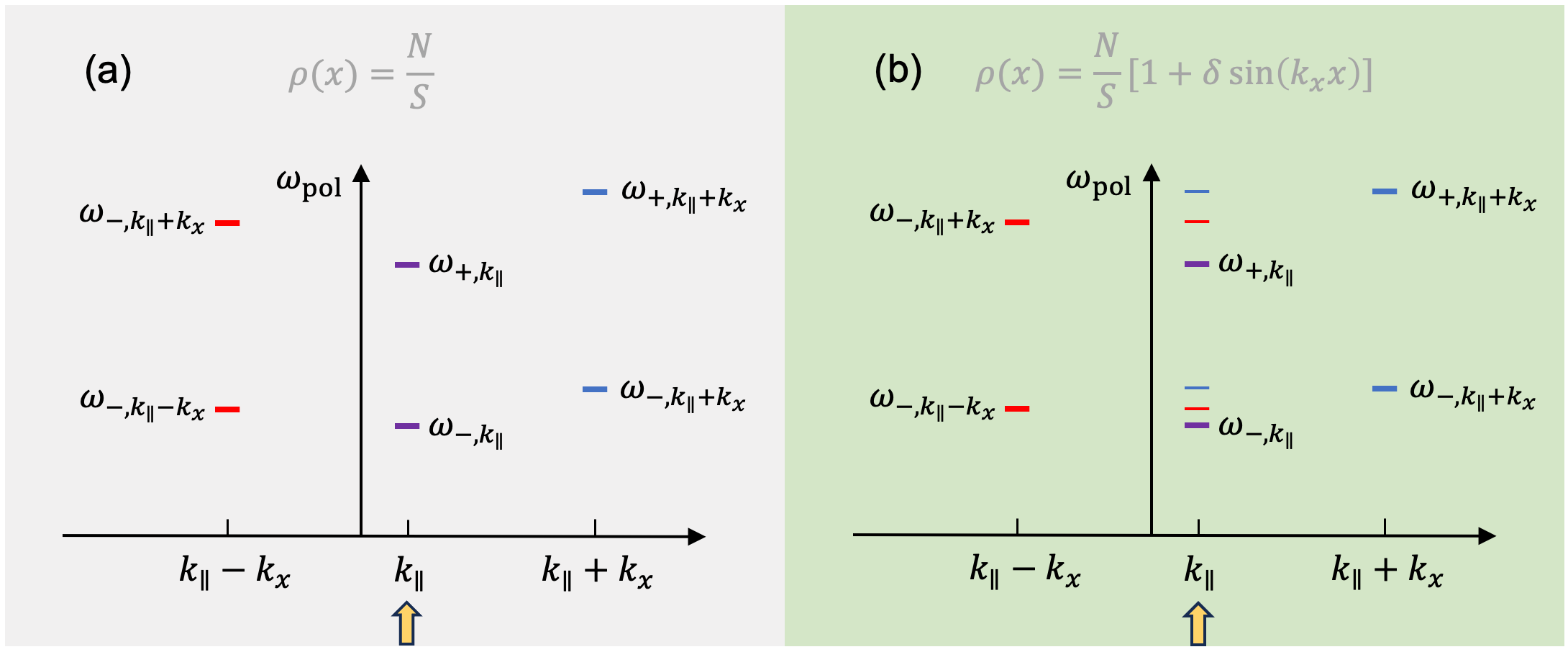}
		\caption{Illustration of the polariton dispersion relations in a planar Fabry--P\'erot cavity under two different cases: (a) a homogeneous molecular density distribution along the 1D cavity mirror plane with $\rho(x) = \frac{N}{S}$; (b) an inhomogeneous molecular density distribution with $\rho(x) = \frac{N}{S}[1 + \delta\sin(k_x x)]$ and $\delta\rightarrow 0$. For part (a), the polariton peaks at only three different in-plane wave vectors, $\kp$ and $\kp\pm k_x$, are explicitly plotted, where $\kp$ is assumed to be near zero. For part (b), \red{the polariton peaks at $\kp$ interact with those at $\kp \pm k_x$, leading to four side peaks at $\kp$. The side polariton peaks at $\kp \pm k_x$ are not shown here.}
		}
		\label{fig:mixed_polariton_states}
	\end{figure}

Eq. \eqref{eq:inhomo_eigen_simplified} can also be rewritten as a set of equations:
\begin{widetext}
\red{
\begin{subequations} \label{eq:eq_inhom_set}
    \begin{align}
    &\cdots \\
    \left [ \omega - \omega_{\rm c}(\kp-k_x) \right ] (\omega - \omega_0)\alpha_{\kp-k_x} & = \Delta^2 \alpha_{\kp-k_x} +  \frac{1}{2i}\delta \Delta^2 \left( \alpha_{\kp-2k_x} - \alpha_{\kp} \right) ,\\
        \left [ \omega - \omega_{\rm c}(\kp) \right ] (\omega - \omega_0)\alpha_{\kp} & = \Delta^2 \alpha_{\kp} +  \frac{1}{2i}\delta \Delta^2 \left( \alpha_{\kp-k_x} - \alpha_{\kp + k_x} \right) , \\
        \left [ \omega - \omega_{\rm c}(\kp + k_x) \right ] (\omega - \omega_0)\alpha_{\kp + k_x} & = \Delta^2 \alpha_{\kp + k_x} +  \frac{1}{2i}\delta \Delta^2 \left( \alpha_{\kp} - \alpha_{\kp + 2k_x} \right) .\\
        & \cdots
    \end{align}
\end{subequations}
}
\end{widetext}
\red{If we are interested in the polariton signals at only $\kp$, because this cavity mode interacts directly with only $\kp \pm k_x$, in Eq. \eqref{eq:eq_inhom_set}} we may discard  the interaction with cavity modes at values \red{greater than $\kp + k_x$  or smaller than $\kp - k_x$, as these cavity modes provide higher-order corrections to the polariton signals at $\kp$.} With this simplification in mind, a closed form is further obtained:
\begin{widetext}
\red{
\begin{subequations}
    \begin{align}
    \label{eq:eHC_pert_0}
   \left\{ \left [ \omega - \omega_{\rm c}(\kp-k_x) \right ] (\omega - \omega_0) - \Delta^2 \right\} \alpha_{\kp-k_x} & \approx  -  \frac{1}{2i}\delta \Delta^2 \alpha_{\kp},\\
    \label{eq:eHC_pert_1}
        \left\{ \left [ \omega - \omega_{\rm c}(\kp) \right ] (\omega - \omega_0) - \Delta^2 \right\} \alpha_{\kp} & =   \frac{1}{2i}\delta \Delta^2 (\alpha_{\kp - k_x}-\alpha_{\kp + k_x}) , \\
        \label{eq:eHC_pert_2}
        \left\{ \left [ \omega - \omega_{\rm c}(\kp + k_x) \right ] (\omega - \omega_0) - \Delta^2 \right\} \alpha_{\kp + k_x} & \approx  \frac{1}{2i} \delta \Delta^2 \alpha_{\kp}  .
    \end{align}
\end{subequations}
}
\end{widetext}
\red{By substituting Eqs. \eqref{eq:eHC_pert_0} and \eqref{eq:eHC_pert_2}, or $\alpha_{\kp \pm k_x} \approx \pm \frac{1}{2i}\delta \Delta^2 \alpha_{\kp} / \left\{ \left [ \omega- \omega_{\rm c}(\kp  \pm k_x) \right ] (\omega - \omega_0)- \Delta^2 \right \}$}, into Eq. \eqref{eq:eHC_pert_1}, we obtain the eigen equation corresponding to $\alpha_{\kp}$:
\red{
\begin{widetext}
\begin{equation}\label{eq:inhomo_four_roots_eq}
    \left\{ \left [ \omega - \omega_{\rm c}(\kp) \right ] (\omega - \omega_0) - \Delta^2 \right \} \left\{ \left [ \omega - \omega_{\rm c}(\kp - k_x) \right ] (\omega - \omega_0) - \Delta^2 \right \} \left\{ \left [ \omega - \omega_{\rm c}(\kp + k_x) \right ] (\omega - \omega_0) - \Delta^2 \right \} = \frac{1}{4}\delta^2 \Delta^4 s(\omega),
\end{equation}
\end{widetext}
where $s(\omega) = \left [ \omega_{\rm c}(\kp + k_x) - \omega_{\rm c}(\kp - k_x) \right ] (\omega - \omega_0)$.
Because $\delta$ is a small number, to the zero-th order approximation, the \red{six} roots of Eq. \eqref{eq:inhomo_four_roots_eq}  are $\omega = \omega_{\pm, \kp}$, $\omega = \omega_{\pm, \kp-k_x}$, and $\omega = \omega_{\pm, \kp+k_x}$. These six roots are the solutions of the following decoupled equations:
\begin{subequations}
    \begin{align}
        \left [ \omega_{\pm, \kp} - \omega_{\rm c}(\kp) \right ] (\omega_{\pm, \kp} - \omega_0) & =  \Delta^2 , \\
                \left [ \omega_{\pm, \kp-k_x} - \omega_{\rm c}(\kp - k_x) \right ] (\omega_{\pm, \kp-k_x} - \omega_0) & = \Delta^2, \\ 
        \left [ \omega_{\pm, \kp+k_x} - \omega_{\rm c}(\kp + k_x) \right ] (\omega_{\pm, \kp+k_x} - \omega_0) & = \Delta^2 .
    \end{align}
\end{subequations}
In other words, $\omega_{\pm, \kp}$ (or $\omega_{\pm, \kp-k_x}$, $\omega_{\pm, \kp+k_x}$) are the polaritonic eigen energies at $\kp$ (or $\kp - k_x$, $\kp + k_x$)  corresponding to a homogeneous molecular distribution, as illustrated in Fig. \ref{fig:mixed_polariton_states}a.
Explicitly, these six eigen energies are given as follows:
\begin{widetext}
\begin{subequations}\label{eq:polariton_energy_perturbative}
    \begin{align}
        \omega_{\pm,\kp} &= \frac{\omega_{0} + \omega_{\rm c}(\kp)}{2} \pm \sqrt{\Delta^2 + \frac{[\omega_0 - \omega_{\rm c}(\kp)]^2}{4}} ,\\
         \omega_{\pm,\kp - k_x} &= \frac{\omega_{0} + \omega_{\rm c}(\kp - k_x)}{2} \pm \sqrt{\Delta^2 + \frac{[\omega_0 - \omega_{\rm c}(\kp - k_x))]^2}{4}} \\
        \omega_{\pm,\kp + k_x} &= \frac{\omega_{0} + \omega_{\rm c}(\kp + k_x)}{2} \pm \sqrt{\Delta^2 + \frac{[\omega_0 - \omega_{\rm c}(\kp + k_x)]^2}{4}}  .
    \end{align}
\end{subequations}
\end{widetext}
}

The above derivations demonstrate that, given a sinusoidal density inhomogeneity in Eq. \eqref{eq:sin_density}, \red{six} polariton peaks emerge at in-plane wave vector $k_{\parallel}$. As illustrated in Fig. \ref{fig:mixed_polariton_states}b, \red{at $k_{\parallel}$, apart from the pair of polariton peaks corresponding to the homogeneous molecular distribution ($\omega_{\pm,k_{\parallel}}$, purple), two additional pairs of polariton peaks ($\omega_{\pm,k_{\parallel}+ k_x}$, blue; $\omega_{\pm,k_{\parallel}- k_x}$, red) appear in the spectra due to the interaction with the cavity modes at $k_{\parallel} \pm k_x$.} Below, at each $k_{\parallel}$ value, the pair of polariton peaks corresponding to the homogeneous molecular distribution will be called as the \textit{main} polariton peaks; the  additional polariton peaks due to the interaction with the other cavity mode(s) will be called as the \textit{side} polariton peaks.

At this point, we have understood that the spatial inhomogeneity of the molecules can cause the polariton states at different $k_{\parallel}$ values to mix with each other. Now, the remaining question is how to quantify the magnitude of this polariton mixing. As in experiments the intensity of the polariton signals is proportional to the photonic contribution in the polaritons, we now calculate the photonic weight of each polariton state mentioned above.

Because we have assumed that the density inhomogeneity is small ($\delta\rightarrow 0$), the photonic weights of the main polariton peaks at $\kp$ should remain mostly the same as the homogeneous limit, i.e., they are the same as those in the JC model:
\begin{subequations}\label{eq:polariton_weight_perturbative}
    \begin{align}
        |\alpha_{\kp}^{\pm, \rm main}|^2 & \approx \frac{(\omega_0 - \omega_{\pm,\kp})^2}{\Delta^2 + (\omega_0 - \omega_{\pm,\kp})^2}.    \end{align}
\red{For the side polaritons due to the interaction with the cavity mode at $\kp - k_x$, the corresponding photonic weights can be obtained using Eq. \eqref{eq:eHC_pert_0}:
    \begin{align}
        |\alpha_{\kp - k_x}^{\pm, \rm side}|^2 & = \frac{\delta^2 |\alpha_{\kp}^{\pm, \rm main}|^2 \Delta^4} {4(\omega_{\pm,\kp} - \omega_{-, \kp-k_x})^2(\omega_{\pm,\kp} - \omega_{+, \kp-k_x})^2}
    \end{align}
}
Similarly, for the side polaritons due to the interaction with the cavity mode at $\kp + k_x$, the corresponding photonic weights can be obtained using Eq. \eqref{eq:eHC_pert_2}:
    \begin{align}
        |\alpha_{\kp + k_x}^{\pm, \rm side}|^2 & = \frac{ \delta^2 |\alpha_{\kp}^{\pm, \rm main}|^2 \Delta^4} {4(\omega_{\pm,\kp} - \omega_{-, \kp+k_x})^2(\omega_{\pm,\kp} - \omega_{+, \kp+k_x})^2}
    \end{align}
\end{subequations}
    The simple analytical forms of the polariton dispersion relation in Eqs. \eqref{eq:polariton_energy_perturbative} and \eqref{eq:polariton_weight_perturbative} quantify the polariton dispersion relation at $\kp$ impacted by a weak sinusoidal density inhomogeneity. 

    \subsubsection{A general perturbative result}
    
    Given an arbitrary weak in-plane density inhomogeneity in 1D, it is possible to express the spatial density distribution as a linear combination of trigonometric functions:
    \begin{equation}\label{eq:arb_density}
    \bar{\rho}(x) = \frac{S}{N}\rho(x)  \approx 1 + \sum_{l=1}^{M} \delta_l \sin(k_l x) + \delta'_l\cos(k_l x) .
    \end{equation}
    where $\bar{\rho}(x)$ denotes the dimensionless molecular density distribution, $\delta_l$ and $\delta'_l$ are small dimensionless variables, and $M$ is a finite number. Then, assuming that the cavity mode at $\kp$ interacts with each of the other cavity modes $\kp \pm k_l$ independently, we can directly use the above results to obtain the polariton dispersion relation corresponding to Eq. \eqref{eq:arb_density}. At the in-plane wave vector $\kp$, apart from the pair of main polariton peaks corresponding to the homogeneous limit, $4M$ side polariton peaks \red{(i.e., $4M$ eigenstates containing non-zero photonic contributions)} may emerge in the spectrum due to the interactions with $2M$ different cavity modes ($\kp \pm k_l$ for $1 \leq l \leq M$). The frequencies of these $4M$ side polariton peaks are the UP and the LP frequencies of the cavity modes at  $\kp \pm k_l$ in the homogeneous limit. At the in-plane wave vector $\kp$, because the density inhomogeneity is weak,  the photonic weights of the main polariton peaks should remain roughly the same as those in the homogeneous limit. For the photonic weights of the side polaritons at $\kp$, if we assume that the side polariton peaks at $\kp$ due to each of the other cavity modes are independent with each other, according to Eq. \eqref{eq:polariton_weight_perturbative}, the photonic weights of the $4M$ side polariton peaks at $\kp$ can be expressed as:
    \begin{subequations}
    \label{eq:photon_weight_pert_arb}
        \begin{align}
        |\alpha_{\kp - k_l}^{\pm, \rm side}|^2 & =  \frac{ (|\delta_l|^2 + |\delta'_l|^2) |\alpha_{\kp}^{\pm, \rm main}|^2 \Delta^4} {4(\omega_{\pm,\kp} - \omega_{-, \kp-k_l})^2(\omega_{\pm,\kp} - \omega_{+, \kp-k_l})^2}, \\
        |\alpha_{\kp + k_l}^{\pm, \rm side}|^2 & =   \frac{ (|\delta_l|^2 + |\delta'_l|^2) |\alpha_{\kp}^{\pm, \rm main}|^2 \Delta^4} {4(\omega_{\pm,\kp} - \omega_{-, \kp+k_l})^2(\omega_{\pm,\kp} - \omega_{+, \kp+k_l})^2} .
    \end{align}
    \end{subequations}
    In the above equations, the first term characterizes the photonic weights of the side polaritons due to the interaction with the main polariton peaks at $\kp - k_l$; the second term characterizes the photonic weights of the side polaritons due to the interaction with the main polariton peaks at $\kp + k_l$. Although our analysis of an arbitrary weak in-plane density inhomogeneity is very preliminary, it demonstrates that the complexity of the polariton dispersion relation directly correlates with the complexity of the in-plane density inhomogeneity in $k$-space.

    If the in-plane density inhomogeneity is large, providing an analytical solution becomes very challenging. Instead, with numerical calculations, we can directly obtain the polariton dispersion relation for an arbitrary in-plane molecular density distribution. In the next section, we will provide numerical details on how to calculate polariton dispersion relations in a brute-force manner. A similar brute-force calculation was performed to study the effects of molecular disorders on the polariton dispersion relations in 1D cavities \cite{Michetti2005}.

    \section{Numerical details}\label{sec:details}
    
    To begin with, the cavity mirror plane was assumed to be 1D (along the $x$-axis).  For an efficient description of the molecules, along the cavity mirror plane, periodic boundary conditions with a cell length of $L_x$ were applied. In each periodic cell, the molecular distribution was modeled by $N^{\rm grid}$ evenly distributed molecules (i.e., harmonic oscillators) along the $x$-axis ranging from $x = 0$ to $L_x$. The density inhomogeneity of molecules was represented by a site-dependent light-matter coupling strength: $g(x) = g_0 \sqrt{\bar{\rho}(x)}$. A larger coupling strength at location $x$ indicates a larger molecular density distribution (or an increase in the thickness of the molecular layer) at this point.  Due to the use of periodic boundary conditions along the cavity mirror plane, the in-plane wave vector  of each cavity mode became discrete: $\kp = \frac{2\pi  l}{L_x}$, where $l = \red{\pm}1, \red{\pm}2, \cdots, \red{\pm}l_x^{\rm max}$ and $l_x^{\rm max}$ denotes the maximal \red{quantum number of the} in-plane cavity modes included in the calculation. The corresponding frequency of each cavity mode was calculated as $\omega_{\rm c}(\kp) = \sqrt{\omega_{\perp}^2 + \omega_{\parallel}^2}$, where $\omega_{\perp}$ denotes the fundamental cavity mode at zero in-plane angle and $\omega_{\parallel} \equiv c|\kp|$.

    Next, for the calculations of the 1D sinusoidal molecular density inhomogeneity, $\bar{\rho}(x) = 1 + \delta\sin(k_x x)$. The following set of parameters was taken to compare the analytical and the numerical results: \red{$\delta = 0.05$}, $k_x = 250$ cm$^{-1}$, $g_0 = 2.0$ cm$^{-1}$, $N^{\rm grid} = 1080$, $l_x^{\rm max} = 100$, $\omega_{\perp} = 2320$ cm$^{-1}$, and $\Delta \kp = \frac{2\pi}{L_x} = 10$ cm$^{-1}$ (or $L_x = 1$ mm). Then, the extended TC Hamiltonian was constructed in a similar form as Eq. \eqref{eq:H_eTC_matrix}, except that the uniform light-matter coupling strength $g_0$ was replaced by the site-dependent coupling strengths $g(x)$. With this set of parameters, the matrix form of the extended TC Hamiltonian had a dimension of \red{$N_{\rm grid} + 2l_x^{\rm max} = 1280$}.  The  Python package \texttt{numpy} was used to diagonalize this matrix. \red{The obtained polariton spectra were numerically converged when  the periodic cell length was set as $L_x = 1$ mm.}

    \red{The situation when the molecular distribution was a periodic 1D Gaussian distribution was also considered. When $0 < x < L_x$, the molecular density distribution obeys:}
    \begin{equation}\label{eq:Gaussian_1d}
        \bar{\rho}(x) = \mathcal{N} \exp\left [{-\frac{(x - L_x/2)^2}{2\sigma^2}}\right ],
    \end{equation}
    \red{and $\bar{\rho}(x) = \bar{\rho}(x + nL_x)$, where $n$ denotes an integer.}
    Here, $L_x = 1$ mm denotes the length of the periodic cells along the cavity mirror plane; the Gaussian width $\sigma$ was chosen as different values; $\mathcal{N}$ denotes a normalization factor which enforces $\avg{\bar{\rho}(x)} \equiv \frac{1}{L_x}\int_{0}^{L_x}dx \bar{\rho}(x) = 1$. All the other parameters were kept the same as the case of the 1D sinusoidal molecular density inhomogeneity.

    Finally, additional calculations were also performed when the cavity mirror plane was assumed to be 2D (along both the $x$- and the $y$-axis). Similar as the 1D calculations, periodic boundary conditions were also applied and the periodic cell had a size of $L_x\times L_y$. $N^{\rm grid}_{x} \times N^{\rm grid}_{y}$ molecules were evenly distributed in this 2D grid.  The same as the 1D cases, the density inhomogeneity of molecules was represented by a  site-dependent light-matter coupling strength: $g(x,y) = g_0 \sqrt{\bar{\rho}(x, y)}$, where   the dimensionless molecular density distribution $\bar{\rho}(x, y) = \frac{S}{N}\rho(x, y)$ will be given later in the manuscript. Along the $x$- or the $y$-axis, the in-plane wave vectors of the cavity modes were discretized: $\kp^x = \frac{2\pi  l}{L_x}$ and $\kp^y = \frac{2\pi  m}{L_y}$,  where \red{$l = \pm 1, \pm 2, \cdots, \pm l_x^{\rm max}$ and $m = \pm 1, \pm 2, \cdots, \pm m_y^{\rm max}$}, respectively. The associated frequency for each cavity mode took the following form: $\omega_{\rm c}(\vkp) = \sqrt{\omega_{\perp}^2 + (\omega_{\parallel}^x)^2 + (\omega_{\parallel}^y)^2}$, where $\omega_{\parallel}^x = c|\kp^x|$ and $\omega_{\parallel}^y = c|\kp^y|$.  The following set of parameters was used: $g_0 = 0.5$ cm$^{-1}$, $N^{\rm grid}_x =  N^{\rm grid}_y = 120$, $l_x^{\rm max} = m_y^{\rm max} = 30$, $\omega_{\perp} = 2320$ cm$^{-1}$, $\Delta \kp^x = \frac{2\pi}{L_x} = 30$ cm$^{-1}$ (or $L_x = 0.33$ mm), and $\Delta \kp^y = \frac{2\pi}{L_y} = 30$ cm$^{-1}$ (or $L_y = 0.33$ mm).
    With this set of parameters, the matrix form of the extended TC Hamiltonian took a dimension of \red{$N_{\rm grid}^x\times N_{\rm grid}^y  + 2l_x^{\rm max}\times 2l_y^{\rm max} = 18000$}. For such a fairly large matrix, instead of \texttt{numpy}, the Python package \texttt{cupy} was used for the matrix diagonalization with a NVIDIA RTX 3090 GPU.

    \section{Results}\label{sec:results}

    \subsection{1D inhomogeneity in the perturbative limit}

    \begin{figure}
		\centering
		\includegraphics[width=1.0\linewidth]{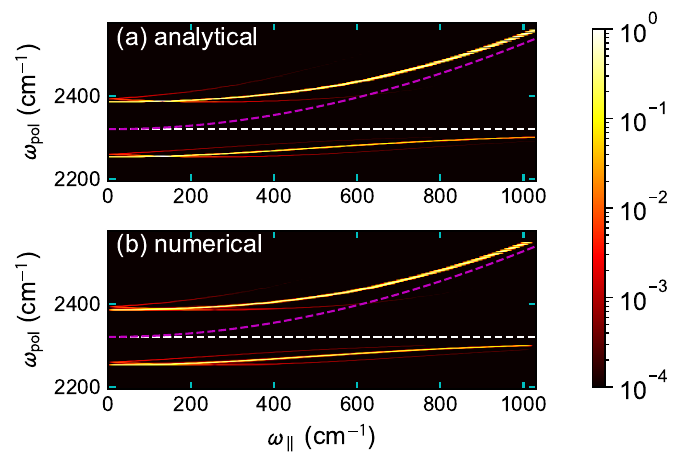}
		\caption{Polariton dispersion relation in a 1D Fabry--P\'erot cavity with a sinusoidal molecular density inhomogeneity: $\bar{\rho}(x) = 1 + \delta\sin(k_x x)$ with \red{$\delta = 0.05$} and $k_x = 250$ cm$^{-1}$. The analytical spectrum [part (a), Eqs. \eqref{eq:polariton_energy_perturbative} and \eqref{eq:polariton_weight_perturbative}] is plotted against the numerical result [part (b)] via directly diagonalizing the extended TC Hamiltonian. For both cases, the polariton intensity (on a logarithmic scale) is represented by the corresponding photonic weight of each eigenstate. The dashed white (magenta) line denote the bare molecular (photonic) excitation energy as a function of the in-plane frequency $\omega_{\parallel} = c|\kp|$.
		}
		\label{fig:1d_analytic}
    \end{figure}
    
    Given the  sinusoidal density inhomogeneity defined in Eq. \eqref{eq:sin_density} [$\bar{\rho}(x) = 1 + \delta\sin(k_x x)$], Fig. \ref{fig:1d_analytic}a plots the polariton dispersion relation using the analytical solution in Eqs. \eqref{eq:polariton_energy_perturbative} and \eqref{eq:polariton_weight_perturbative} when the small parameter $\delta$ is chosen as \red{$\delta = 0.05$}.  In Fig. \ref{fig:1d_analytic}a, at each individual in-plane cavity frequency $\omega_{\parallel}$ ($=c|k_{\parallel}|$), apart from the pair of main polariton peaks, \red{four}  side polariton peaks with weaker photonic weights appear in the spectrum\red{, in agreement with our analysis around Fig. \ref{fig:mixed_polariton_states}b.}
        
    \begin{figure}
		\centering
		\includegraphics[width=1.0\linewidth]{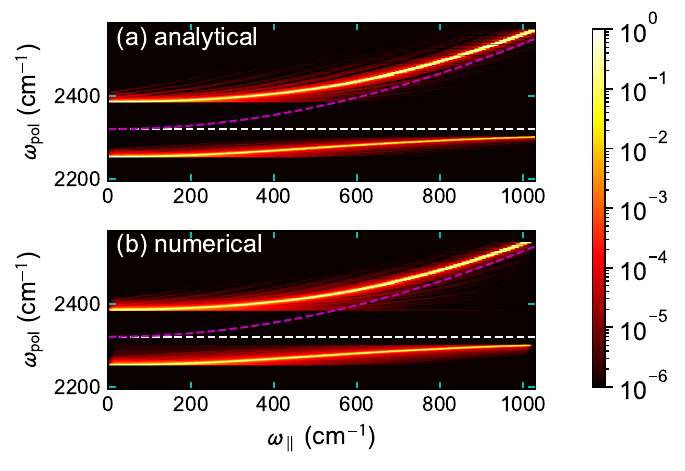}
		\caption{Polariton dispersion relation in a 1D Fabry--P\'erot cavity with the   molecular density inhomogeneity defined in Eq. \eqref{eq:sin_inhom_general}. The analytical spectrum [part (a), Eqs. \eqref{eq:photon_weight_pert_arb}] is plotted against the numerical result [part (b)] via directly diagonalizing the extended TC Hamiltonian.
		}
		\label{fig:1d_analytic_general}
    \end{figure}
    
    To further examine our derivations, we perform numerical calculations of the polariton dispersion relation by directly diagonalizing the extended TC Hamiltonian. Fig. \ref{fig:1d_analytic}b plots the corresponding numerical polariton spectrum. 
    The agreement between Fig. \ref{fig:1d_analytic}a and Fig. \ref{fig:1d_analytic}b cross-validate both our analytical derivations and the numerical calculations.

    Fig. \ref{fig:1d_analytic_general} provides the comparison between the analytical and the numerical polariton dispersion relation for a more complicated weak 1D inhomogeneity:
    \begin{equation}\label{eq:sin_inhom_general}
        \bar{\rho}(x) = 1 +  \frac{\delta}{N_{\rm p}} \sum_{n=1}^{N_{\rm p}} \sin(k_n x).
    \end{equation}
    Here, the small parameter is chosen as  $\delta = 0.1$, $k_n = 10p_n$ cm$^{-1}$, and $p_n = 2, 3, 5, \cdots, 43$ denote $N_{\rm p} = 13$ prime numbers. As a generalization of Fig. \ref{fig:1d_analytic}, both analytical results [using Eq. \eqref{eq:photon_weight_pert_arb}] and numerical calculations demonstrate a comb of weak side polariton branches near the main upper and lower polariton branches. The consistency in Figs. \ref{fig:1d_analytic_general}a,b  provides another cross-validation between the analytical and the numerical calculations.

    \begin{figure}
		\centering
		\includegraphics[width=1.0\linewidth]{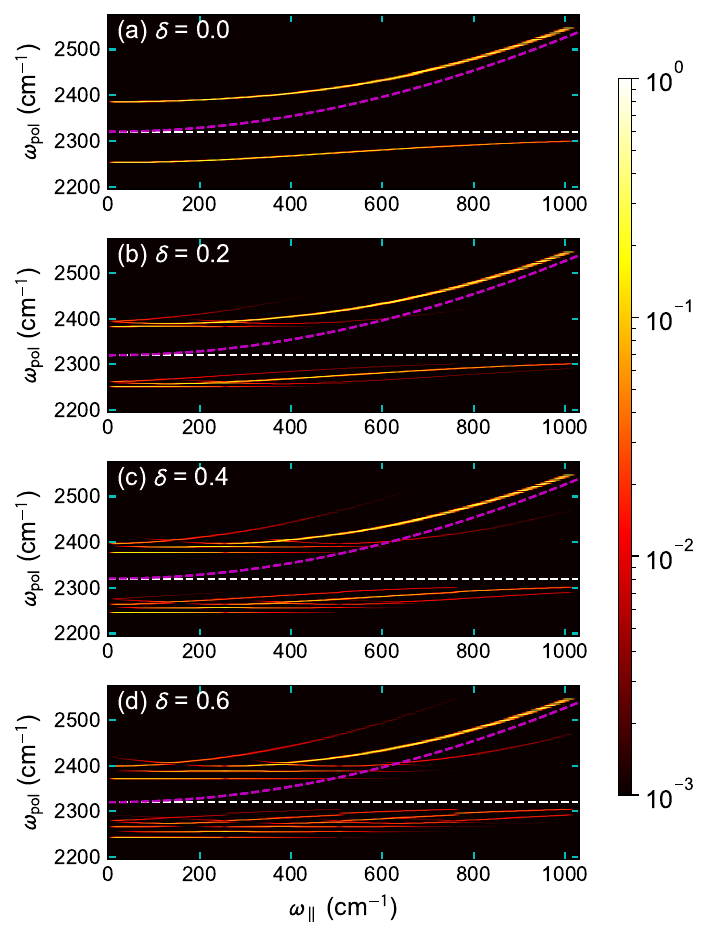}
		\caption{Numerically calculated polariton dispersion relation corresponding to the  1D sinusoidal molecular density inhomogeneity: $\bar{\rho}(x) = 1 + \delta\sin(k_x x)$. In each subplot,  $k_x = 250$ cm$^{-1}$ and the inhomogeneity amplitude is tuned to (a) $\delta = 0.0$ (the homogeneous limit), \red{(b) $\delta = 0.2$,  (c) $\delta = 0.4$, and (d) $\delta = 0.6$}.
		}
		\label{fig:1d_amp}
    \end{figure}

    \subsection{1D sinusoidal inhomogeneity beyond the perturbative limit}

    Moving forward, using the sinusoidal molecular density inhomogeneity $\bar{\rho}(x) = 1 + \delta\sin(k_x x)$, we perform additional numerical calculations to investigate the modification of polariton dispersion relations beyond the perturbative limit. Fig. \ref{fig:1d_amp} plots the polariton spectra with different values of $\delta$, the amplitude of the sinusoidal density inhomogeneity. When $\delta = 0$ (Fig. \ref{fig:1d_amp}a, the homogeneous limit), only two polariton branches are obtained,  corresponding to the conventional polariton dispersion relation in the homogeneous limit. When \red{$\delta = 0.2$, 0.4, and 0.6} (Figs. \ref{fig:1d_amp}b-d), apart from the two main polariton branches, more and more side polariton branches appear in the spectra. The number increase of the side polariton branches indicate the higher-order interactions between different cavity modes, which are completely discarded in our analytical derivations. Very interestingly, in these spectra, different side polariton branches are disconnected with a spacing of 250  cm$^{-1}$, indicating the interaction between cavity modes at  $\kp$ and \red{$\kp \pm k_x$}, as $k_x = 250$ cm$^{-1}$ was kept the same during the calculations.  When \red{$\delta = 0.6$} (Fig. \ref{fig:1d_amp}d),  the main polariton branches are strongly altered by the side polariton branches. As a result, in this strong inhomogeneity limit, the definition of the main branches becomes obscure.

    \begin{figure}
		\centering
		\includegraphics[width=1.0\linewidth]{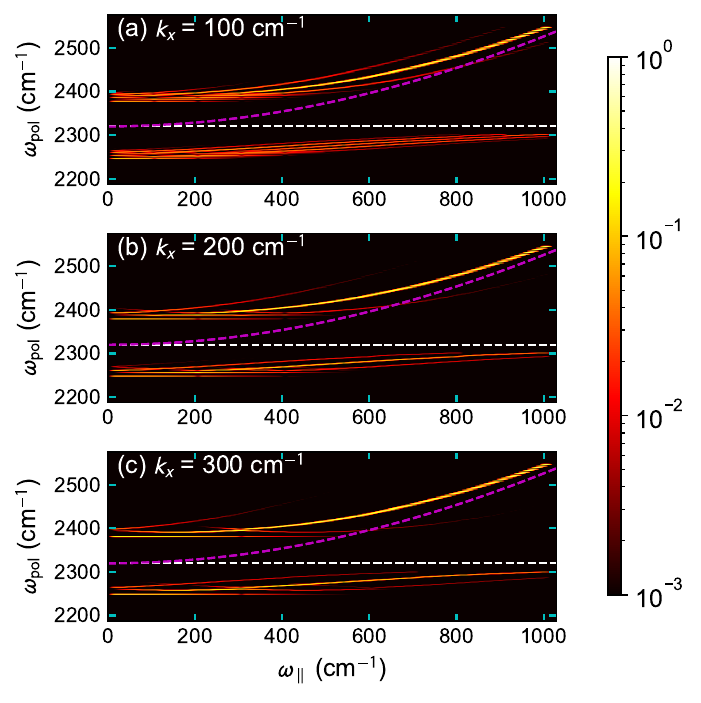}
		\caption{Numerically calculated polariton dispersion relation corresponding to the 1D sinusoidal molecular density inhomogeneity: $\bar{\rho}(x) = 1 + \delta\sin(k_x x)$. In each subplot,  the inhomogeneity amplitude \red{$\delta = 0.3$}, and $k_x$ is tuned to (a) 100 cm$^{-1}$, (b) 200 cm$^{-1}$, and (c) 300 cm$^{-1}$.
		}
		\label{fig:1d_freq}
    \end{figure}

    Fig. \ref{fig:1d_freq} plots a series of polariton spectra when $\delta  = 0.3$ is fixed and $k_x$ is tuned to (a) 100 cm$^{-1}$, (b) 200 cm$^{-1}$, and (c) 300  cm$^{-1}$, respectively. Because here $\delta$ is relatively large, multiple side polariton branches exist in each spectrum. \red{As $k_x$ is reduced, the side polariton branches become more dense.} However, in each spectrum, different side polariton branches are still disconnected with a spacing of the corresponding $k_x$ value. 

    \subsection{1D Gaussian distribution} 
    
    \begin{figure}
		\centering
		\includegraphics[width=1.0\linewidth]{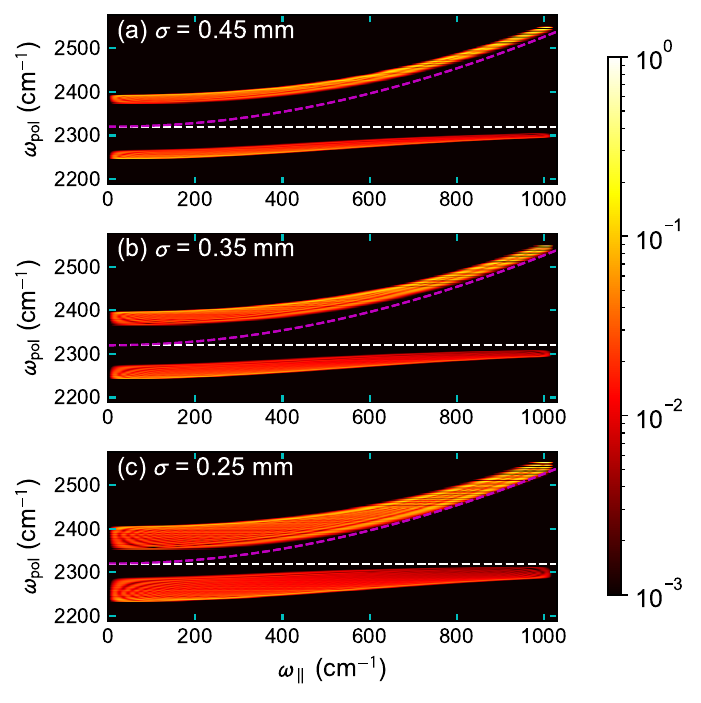}
		\caption{Numerically calculated polariton dispersion relation corresponding to the 1D \red{periodic} Gaussian molecular in-plane density distribution defined in Eq. \eqref{eq:Gaussian_1d}. In each subplot,  the Gaussian width $\sigma$ is tuned to (a) $0.45$ mm, (b) $0.35$ mm, and (c) $0.25$ mm. Reducing the Gaussian width, i.e., increasing the molecular distribution inhomogeneity, enhances the broadening of the two main polariton branches.
		}
		\label{fig:1d_gaussian}
    \end{figure}

    Fig. \ref{fig:1d_gaussian} plots the polariton dispersion relation when the molecular distribution obeys the  1D \red{periodic} Gaussian distribution defined in Eq. \eqref{eq:Gaussian_1d}.
    When the width of the Gaussian distribution is $\sigma = 0.45$ mm, the corresponding polariton dispersion relation (Fig. \ref{fig:1d_gaussian}a) is very different from the cases of the  sinusoidal inhomogeneity. Here, instead of the appearance of a few discrete side polariton branches, only the two main polariton branches remain. However, each polariton branch becomes significantly broadened. This polariton broadening can be understood as follows: as discussed around Eq. \eqref{eq:photon_weight_pert_arb}, because the Fourier transform of a spatial Gaussian distribution is still a Gaussian distribution in the frequency domain, the cavity mode at $k_{\parallel}$ can interact with all the cavity modes within the frequency neighbourhood \red{$[k_{\parallel} - \Delta k, k_{\parallel} + \Delta k]$}, where $\Delta k \propto 1/\sigma$. As a result, instead of the emergence of a few discrete side polariton branches, an enormous number of side polariton branches appear in the frequency neighbourhood of $\kp$. Hence,  the original two main polariton branches become effectively broadened. In Fig. \ref{fig:1d_gaussian}b,c, when the Gaussian distribution has a smaller width ($\sigma = 0.35$ mm or 0.25 mm), i.e., when the molecular distribution becomes more inhomogeneous, the polariton broadening appears to be more significant. This trend agrees with our analysis above that the linewidth broadening at $k_{\parallel}$  is due to the interaction with the cavity modes \red{within the range $[k_{\parallel} - \Delta k, k_{\parallel} + \Delta k]$}.
    
    While it is known that the polariton linewidth broadening can be attributed to the homogeneous broadening of the molecular linewidth \cite{Houdre1996,Long2015}, the cavity loss, \red{or the strong energetic disorder of molecules}, the intriguing polariton broadening effect in Fig. \ref{fig:1d_gaussian} clearly shows that the large-scale \red{(in the order of $\sigma \sim 0.1$ mm)}, in-plane molecular density inhomogeneity can perhaps increase the polariton linewidth in a very significant manner. Such a linewidth broadening comes from the breakdown of the single-mode approximation and the emergence of the side polariton peaks due to the in-plane translational symmetry breaking. \red{This finding provides another perspective to understand the origins of the polariton linewidths observed in various VSC experiments \cite{Thomas2016,Thomas2019_science,Ahn2023Science,Imperatore2021,Wiesehan2021,Fidler2023,Hirai2020Crys}, where the chemical or phase transition processes may potentially prohibit the molecular system from maintaining the in-plane homogeneity on such a large length scale ($\sigma \sim 0.1$ mm). This possibility remains exploration in the future. Note that when $\sigma$ is small enough, the resulting polariton linewidth may be comparable with or large than the Rabi splitting. Under this extreme inhomogeneous limit, the system would transit from strong coupling to weak coupling.}

    \subsection{2D distributions} 

    The above calculations assumed that the cavity mirror plane was only 1D. This assumption might be questionable when modeling planar Fabry--P\'erot cavities with 2D cavity mirrors. To this end, we perform additional calculations to directly study the polariton dispersion relation when the cavity mirror plane becomes 2D.

    \begin{figure}
		\centering
		\includegraphics[width=1.0\linewidth]{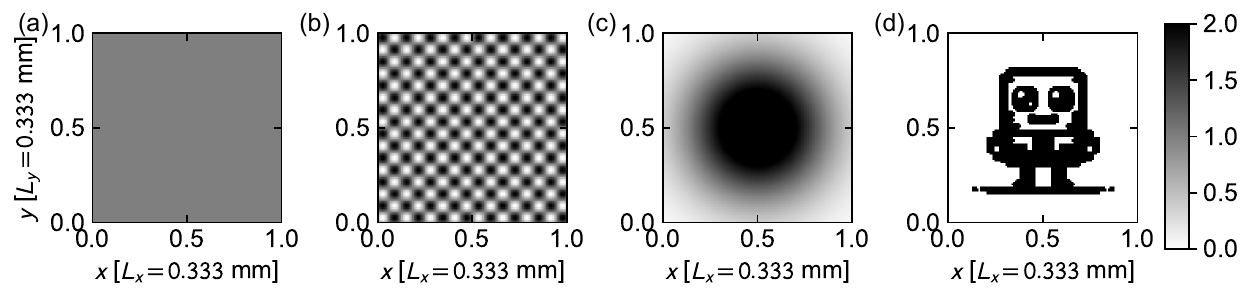}
		\caption{The four molecular density distributions $\bar{\rho}(x,y)$ considered in the 2D calculations: (a) the homogeneous limit when $\bar{\rho}(x,y) = 1$; (b) the 2D sinusoidal inhomogeneity [Eq. \eqref{eq:2D_distr_sin}]; (c) the 2D Gaussian distribution [Eq. \eqref{eq:2D_distr_Gaussian}]; (d) a cartoon pattern. For each case, periodic boundary conditions are applied along the $xy$-plane, and the density distribution in a single periodic cell is shown.
		}
		\label{fig:molecular_dist_2d}
    \end{figure}

\begin{figure}
		\centering
		\includegraphics[width=1.0\linewidth]{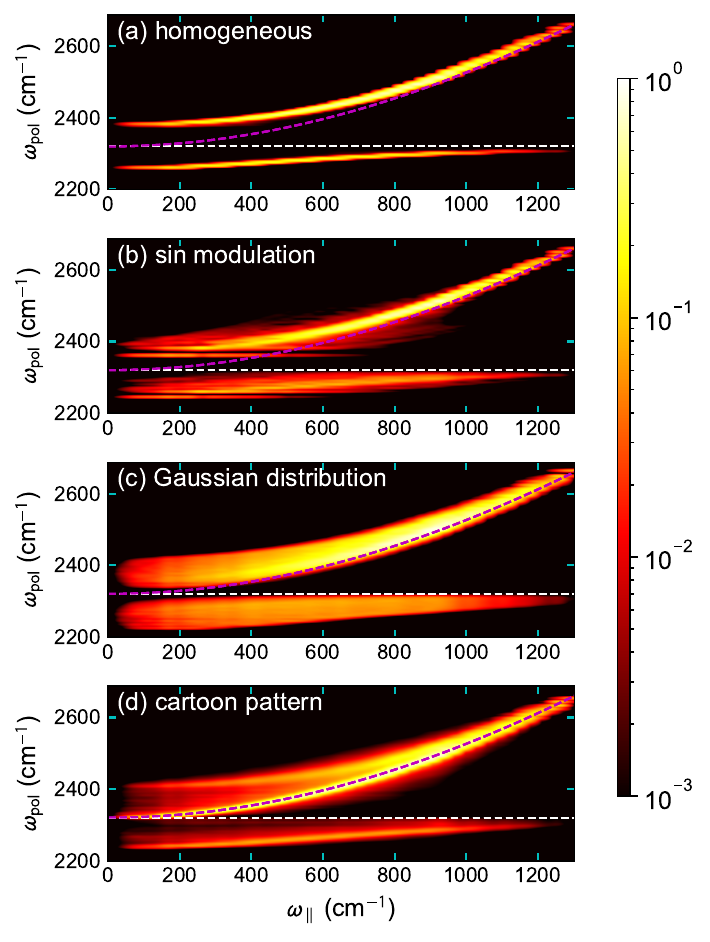}
		\caption{The polariton dispersion relations corresponding to the four 2D molecular distributions in Fig. \ref{fig:molecular_dist_2d}: \red{(a) the homogeneous limit; (b) the 2D sinusoidal inhomogeneity; (c) the 2D Gaussian distribution; (d) the cartoon pattern. }
		}
		\label{fig:2d_compare}
    \end{figure}

    As shown in Fig. \ref{fig:molecular_dist_2d}, four different 2D molecular distributions are considered. For the first homogeneous case (Fig. \ref{fig:molecular_dist_2d}a), the corresponding polariton spectrum is plotted in Fig. \ref{fig:2d_compare}a. Here, only two polariton branches are obtained, in agreement with the 1D case (Fig. \ref{fig:1d_amp}a). Although the spectrum is similar as the 1D case, a significant difference is needed to be emphasized: in the 1D calculations, the in-plane frequency $\omega_{\parallel}$ refers to $\omega_{\parallel}^x$, the cavity in-plane frequency along the cavity mirror plane direction (the $x$-direction); in the 2D calculations, the in-plane frequency $\omega_{\parallel}$ refers to $\sqrt{|\omega_{\parallel}^x|^2 + |\omega_{\parallel}^y|^2}$, an arbitrary combination of the in-plane frequencies in both the $x$- and the $y$-direction. In this homogeneous case, due to the preservation of the rotational symmetry along the 2D plane,  $\omega_{\parallel} = \sqrt{|\omega_{\parallel}^x|^2 + |\omega_{\parallel}^y|^2}$ is a good quantum number to characterize the cavity frequency dependence of the polariton spectrum.

    Then, we consider a 2D sinusoidal molecular density inhomogeneity  (Fig. \ref{fig:molecular_dist_2d}b):
    \begin{equation}\label{eq:2D_distr_sin}
        \bar{\rho}(x, y) = 1 + \delta \sin(k_x x)\sin(k_y y)
    \end{equation}
    where $\delta = 1.0$ and $k_x = k_y = 250$ cm$^{-1}$. 
    
    Fig. \ref{fig:2d_compare}b plots the  polariton spectrum corresponding to the above 2D sinusoidal molecular density inhomogeneity. Here, apart from the two main polariton branches, a few side polariton branches can still be observed. Compared to the 1D correspondence, in an effort to efficiently calculate the 2D results, we have used a larger frequency spacing between adjacent cavity modes per dimension\red{; see Sec. \ref{sec:details} for details}. As a result, the frequency resolution here is relatively low. Nevertheless, the existence of a few side polariton branches suggests that the observation in the 1D sinusoidal distributions can still be valid in 2D.
    
    Next, repeating the 1D calculations, we also consider a 2D periodic Gaussian density distribution (Fig. \ref{fig:molecular_dist_2d}c). \red{When $0 < x < L_x$ and $0 < y< L_y$, the Gaussian distribution obeys:}
    \begin{equation}\label{eq:2D_distr_Gaussian}
         \bar{\rho}(x, y) = \mathcal{N} \exp\left [{-\frac{(x - \frac{L_x}{2})^2 + (y - \frac{L_y}{2})^2}{2\sigma^2}}\right ],
    \end{equation}
    \red{and $\bar{\rho}(x, y) = \bar{\rho}(x + nL_x, y)$ and $\bar{\rho}(x, y) = \bar{\rho}(x, y + nL_y)$, where $n$ denotes an integer. For parameters,} $L_x = L_y = $ 0.333 mm, $\sigma = 0.083$ mm, and $\mathcal{N}$ denotes a renormalization factor which enforces $\avg{\bar{\rho}(x, y)} \equiv \frac{1}{L_xL_y}\int_{0}^{L_x}dx \int_{0}^{L_y}dy \bar{\rho}(x, y) = 1$. For this 2D Gaussian distribution, Fig. \ref{fig:2d_compare}c plots the corresponding polariton dispersion relation. Here, two broadened polariton branches are observed, in  agreement with the 1D correspondence (Fig. \ref{fig:1d_gaussian}). 
    
    Finally, we perform an additional calculation when the molecular density distribution becomes Fig. \ref{fig:molecular_dist_2d}d, a cartoon pattern \red{which lacks symmetry}. For such a density distribution, the corresponding polariton spectrum (Fig. \ref{fig:2d_compare}d) demonstrates three major peaks: the two polariton branches as in Fig. \ref{fig:2d_compare}a, and a purely photonic excitation (the magenta line). The purely photonic excitation is related to the fact that there is some empty space in the 2D cartoon distribution. As a result, the photonic modes can sometimes be completely decoupled from the molecular excitations. 
    
    As a side note, because the molecular density distributions \red{in Figs. \ref{fig:molecular_dist_2d}b,d} do not preserve the in-plane rotational symmetry,  the in-plane frequency $\omega_{\parallel} = \sqrt{|\omega_{\parallel}^x|^2 + |\omega_{\parallel}^y|^2}$ may not be a good quantum number to describe the polariton spectrum. Instead, a better solution is to plot the polariton spectrum as a function of both $\omega_{\parallel}^x$ and $\omega_{\parallel}^y$, i.e., plotting a 3D polariton dispersion relation \cite{doi:10.1126/sciadv.abq7533}.

    \section{Discussion}\label{sec:discussion}

    In an effort to better understand how \red{the cavity modes at different $\kp$ values} can mix with each other when the in-plane molecular density distribution becomes inhomogeneous, we further study the photonic weight distribution among the eigenstates in the VSC systems with broken in-plane translational symmetry. 

    \begin{figure}
		\centering
		\includegraphics[width=1.0\linewidth]{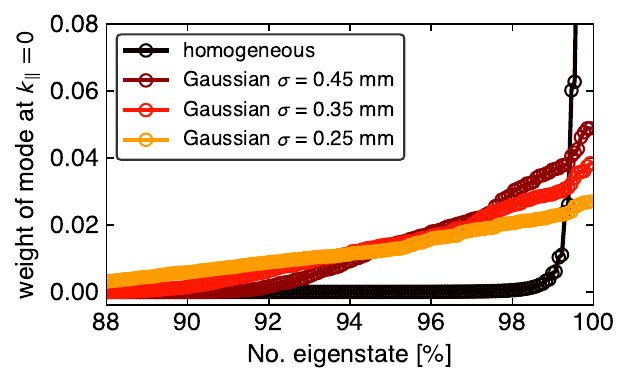}
		\caption{The $\kp=0$  photonic weight distribution among all VSC eigenstates under a few different 1D conditions: the homogeneous limit (black) is compared against the three Gaussian distributions in Fig. \ref{fig:1d_gaussian}: $\sigma = 0.45$ mm (brown), $\sigma =$ 0.35 mm (red), and $\sigma =$ 0.25 mm (orange). The photon weights of different eigenstates are plotted in the ascend order. 
        More eigenstates contain a non-zero $\kp=0$ cavity photon contribution when the molecular in-plane distribution becomes more inhomogeneous. 
		}
		\label{fig:distr_eigen}
    \end{figure}

    Fig. \ref{fig:distr_eigen} quantifies the weight distribution of the $\kp = 0$ cavity photon mode among the VSC eigenstates in a few different 1D systems. \red{Because our numerical calculations use discretized $\kp$ ($= 2\pi l /L_x$), the $\kp = 0$ cavity photon mode refers to the case when $l=\pm 1$. } The homogeneous limit (black) is compared against the three Gaussian distributions discussed in Fig. \ref{fig:1d_gaussian}: the distributions with Gaussian widths $\sigma = 0.45$ mm (brown), $\sigma = 0.35$ mm (red), and $\sigma = 0.25$ mm (orange). For each case, the photon weights of different eigenstates are ranked in the ascend order.
    In the homogeneous limit, more than 99\%  of the eigenstates contain zero $\kp = 0$ photonic contribution, demonstrating a sharp separation between the bright (or polaritonic) and the dark modes. 
    For results corresponding to the Gaussian distributions, when the  Gaussian width is reduced (or with an increased density inhomogeneity), more eigenstates contain a non-zero contribution of the $\kp = 0$ cavity photon mode. This trend agrees with the increased polariton linewidths (i.e., more side polariton peaks) as observed in Fig. \ref{fig:1d_gaussian}. 
    Especially, when $\sigma = 0.25$ mm, \red{more than 10\%} of the eigenstates contain a non-zero $\kp = 0$ photonic contribution.

    Previous experiments have indeed indicated the possibility of preparing the so-called gray states, i.e., a large number of dark modes containing a non-zero photonic contribution, using  materials with large energy disorder \cite{Son2022,George2024}.  However, our manuscript has demonstrated a very universal engineering technique, i.e., changing the molecular in-plane density distribution without altering the intrinsic properties of molecules, to achieve similar goals with the mechanism of in-plane translational symmetry breaking.

    \section{Conclusion}\label{sec:conclusion}

    In summary, we have studied vibrational polaritons with patterned in-plane molecular density distributions.  Due to the presence of the molecular density inhomogeneity (or the lack of in-plane translational symmetry), the single-mode approximation usually invoked in planar Fabry--P\'erot microcavities becomes invalid, and the scattering between cavity modes at different in-plane wave vectors  must be taken into account. As a result, even in the case of  uniform molecular excitation frequencies plus simple planar Fabry--P\'erot geometries (which have been assumed throughout this manuscript), complicated polariton dispersion relations could emerge by tuning the molecular density distribution along the cavity mirror plane. 
    
    In detail, for the case of a weak sinusoidal modulation of the homogeneous in-plane molecular density distribution, 1D perturbative calculations suggest that, in addition to a pair of main polariton branches, four additional side polariton branches could appear in the polariton dispersion relation. Numerical calculations further suggest the number increase of the side polariton branches when the sinusoidal inhomogeneity becomes stronger. More interestingly, for the case of a \red{periodic} Gaussian molecular density distribution, numerical calculations demonstrate that the two main polariton branches can become significantly broadened. Such a polariton linewidth broadening is distinguished from other well-studied origins of polariton broadening\red{\cite{Houdre1996,Long2015,George2024}}: here the broadening results from \red{the large-scale ($\sigma\sim 0.1$ mm) in-plane density inhomogeneity and the significant interactions between cavity modes at different $\vkp$  values}.

    While our derivations and calculations have used the parameters for VSC, the conclusions of this manuscript can also be applied to some other types of collective strong couplings in planar Fabry--P\'erot cavities such as exciton-polaritons. However, we believe that VSC might be a more advantageous platform for an experimental verification of our conclusions than electronic strong coupling. \red{This is because, due to the cavity frequency difference,} the fabrication of in-plane molecular density inhomogeneity for VSC requires a spatial resolution of $0.01 \sim 0.1$ mm (as shown in Fig. \ref{fig:molecular_dist_2d}), which should be one-order-of-magnitude larger than that of electronic strong coupling.

    From a theoretical perspective, we have further emphasized the blurring between the bright and the dark modes with a \red{periodic} Gaussian molecular in-plane density distribution. Whether such a state blurring could impact polariton transport or even molecular dynamics in the dark is worthy of further investigation.

    \section{Acknowledgments}
    This work is supported by  start-up funds from the University of Delaware Department of Physics and Astronomy. 

    \section{Data Availability Statement}
    \red{The code and data that support the findings of this study are available at Github: \url{https://github.com/TaoELi/vsc_symmetry_breaking}.}


    %

    \end{document}